\documentclass{mn2e}
\usepackage{graphicx,times}

\def \apjs{ApJS}

\def \apj{ApJ}

\def \mnras{MNRAS}

\def \mic{$\mu$m}
\def \kms{km\,s$^{-1}$}
\def \nai{Na\,{\sc i}}

\def \naix{Na\,{\sc i} 1.14\,$\mu$m}
\def \naiop{Na\,{\sc i} 0.82\,$\mu$m}

\def \rev  {}

\title[The 1.14-\mic\ Na\,{\sc i} doublet]{The IMF-sensitive 1.14-\mic\ \nai\ doublet in early-type galaxies}

\author[Russell J. Smith et al.]{Russell J. Smith$^1$, 
Padraig Alton$^1$, John R. Lucey$^1$, Charlie Conroy$^2$ and David Carter$^3$\\
$^1$ Centre for Extragalactic Astronomy, Department of Physics, Durham University, South Road, Durham \ DH1 3LE \\
$^2$ Harvard Smithsonian Centre for Astrophysics, 60 Garden St., Cambridge, MA 02138, USA\\
$^3$ Astrophysics Research Institute, Liverpool John Moores University, IC2, Liverpool Science Park 146 Brownlow Hill, Liverpool L3 5RF
}

\pagerange{\pageref{firstpage}$-$\pageref{lastpage}} \pubyear{2014}

\begin{document}

\date{Accepted 2015 September 02. Received 2015 July 31; in original form 2015 May 13.}

\label{firstpage}

\voffset=-1.5cm

\maketitle

\begin{abstract}
We present J-band spectroscopy of passive galaxies focusing on the Na\,{\sc i} doublet at 1.14\,\mic.
Like the \naiop\ doublet, this feature is strong in low-mass stars and hence may provide a useful probe of the initial mass function (IMF).
From high signal-to-noise composite spectra, we find that \naix\ increases steeply with increasing velocity dispersion, $\sigma$, 
and for the most massive galaxies ($\sigma$\,$\ga$\,300\,\kms) is much stronger than predicted from synthetic spectra with Milky-Way-like IMFs and solar abundances. 
Reproducing \naix\ at high-$\sigma$ likely requires either a very high
[Na/H], or a bottom-heavy IMF, 
or a combination of both. 
Using the Na\,D line to break the degeneracy between IMF and abundance, we infer [Na/H]\,$\approx$\,+0.5 and a steep IMF (single-slope-equivalent $x$\,$\approx$\,3.2,
where $x$\,=\,2.35 for Salpeter), for the  high-$\sigma$ galaxies. At lower mass ($\sigma$\,=\,50--100\,\kms), the line strengths are compatible with MW-like IMFs and near-solar [Na/H].
We highlight two galaxies in our sample where strong gravitational lensing masses favour MW-like IMFs. 
Like the high-$\sigma$ sample on average, these galaxies have strong \naix; taken in isolation their sodium indices imply bottom-heavy IMFs which are
hard to reconcile with the lensing masses.
An alternative full-spectrum-fitting approach, applied to the high-$\sigma$ sample, recovers an IMF less heavy than Salpeter, but under-predicts
the \naix\ line at the 5$\sigma$ level.
We conclude that current models struggle to reproduce this feature in the most massive galaxies
without breaking other constraints, and caution against over-reliance on the sodium lines in spectroscopic IMF studies.

\end{abstract}

\section{Introduction}

Because the luminosity-vs-mass relation for stars is steeper than the distribution function of their masses, dwarf stars contribute very little to the integrated light from an old stellar population.
However it has long been realised that some gravity-sensitive spectral features, which become extremely strong in late M dwarfs, provide a route to constraining the low-mass 
stellar content, and hence the initial mass function (IMF), in unresolved galaxies 
(e.g. Spinrad \& Taylor 1971; Whitford 1977; 
Carter, Visvanathan \& Pickles 1986).
The \nai\ doublet absorption at 0.82\,\mic\ is among the strongest and most readily accessible of these dwarf-star indicators.

Recent work has advanced this method using measurements with red-sensitive CCDs, large spectroscopic surveys, and improved near-infra-red (IR) instrumentation
(e.g. van Dokkum \& Conroy 2010; Spiniello et al. 2012; Conroy \& van Dokkum 2012b; Smith, Lucey \& Carter 2012; La Barbera et al. 2013). In parallel, more 
extensive stellar libraries (e.g. Rayner et al. 2008) and sophisticated spectral synthesis models (e.g. Conroy \& van Dokkum 2012a, CvD12a; Vazdekis et al. 2012) 
have led to better quantification of the IMF sensitivity of various features, and the degeneracies with element abundance variations.

These studies found that the  \naiop\ doublet observed in massive
early-type galaxies  is much stronger than would be expected from an old stellar population with an IMF like that in the 
Milky Way, and with solar element abundances (van Dokkum \& Conroy 2010; %
Spiniello et al. 2012; Conroy \& van Dokkum 2012b; Fererras et al. 2013).
The observations can be reproduced by adopting either a bottom-heavy IMF or a highly-enhanced sodium abundance, or
a combination of both. As an example, for NGC\,4621, the full-spectrum fitting analysis of Conroy \& van Dokkum (2012b) requires an IMF with a
mass-to-light ratio larger by a factor $\alpha$\,=\,1.95$\pm$0.05, compared to the Milky Way IMF, and a sodium abundance [Na/H]\,=\,+0.84$\pm$0.04 (their figure 3). 
However, the results are sensitive to the spectral regions used in the fit: 
if \naiop\ is excluded, they derive $\alpha$\,$\approx$\,1.2 for this galaxy, not much in excess of a Kroupa (2001) IMF ($\alpha$\,=\,1), while if 
the Ca\,{\sc ii} triplet is {\it also} excluded, a heavy IMF is again recovered (their figure 12).

Since \naiop\ plays an important role in the evidence for non-standard IMFs in massive galaxies, 
and because alternative model sets disagree in {\it predicting} the \naiop\ and Na\,D line behaviour (Spiniello, Trager \& Koopmans 2015), it is worthwhile not only 
to explore other indicators based on different atomic/molecular species (e.g. Spiniello et al. 2014), but also to test models against a wider set of observed sodium features.

In this {\it Letter}, we focus on the J-band \nai\ doublet absorption at 1.14\,$\mu$m. This line arises from atoms 
in the same energy level as \naiop, and hence should have similar ``fundamental'' temperature and pressure dependence,
although blending with other features inevitably affects the behaviours differently in practice (e.g., \naiop\ coincides with a strong Ti\,O band).
As expected, \naix\ is very strong in cool dwarf stars, and weak in giants (Rayner et al. 2008), yielding good predicted sensitivity to the IMF (Conroy \& van Dokkum 2012a). 
No published work has yet analysed \naix\ in early-type galaxies, most IR studies to date having been made at longer wavelengths
(Silva, Kuntschner \& Lyubenova 2008; Marmol-Queralt\'o et al. 2009; Cesetti et al. 2009). 
For very nearby galaxies, accurate measurement of \naix\ is hampered by telluric water absorption at 1.11-1.16\,\mic.
Here, we study two samples of galaxies at modest distance ($z$\,=\,0.02--0.06), so that \naix\ is redshifted into a cleaner atmospheric window.
We use rest-frame stacking to build composite spectra, further suppressing the sky and telluric noise contributions.

{\rev As this is the first analysis of the \naix\ feature in this context, our treatment is exploratory in nature, and intended to capture the ``first order''
behaviour and to distinguish IMF from sodium abundance effects, rather than to provide an exhaustive treatment of all possible parameter degeneracies.}

Our samples, observations and data treatment are described in Section~\ref{sec:data} and the results are presented in Section~\ref{sec:nai}.
We summarize our conclusions in Section~\ref{sec:disc}.

\section{Data}\label{sec:data}

We combine  J-band data from two observational programmes to measure \naix\ over a wide range in galaxy mass, and  derive Na\,D from 
optical survey spectroscopy for the same galaxies.

\subsection{Subaru Coma sample}

We obtained J-band spectra for 84 red-sequence galaxies in the Coma cluster, with velocity dispersions $\sigma$\,=\,50--300\,\kms,
using FMOS, the IR Fiber Multi-Object Spectrograph at the 8.2m Subaru Telescope (Kimura et al. 2010).
The observations were made on the nights of 2014 April 9--10,  in the ``J-long'' configuration, with spectral resolution $R$\,$\approx$\,2400.
The target galaxies, fibre configurations and observing strategy were the same as those used in the Y-band (``J-short'') observations reported by Smith et al. (2012b).
Four fibre configurations were observed in cross-beam-switching mode, with fainter galaxies repeated on more configurations, for a total integration of up to 10.5\,h per target.

Initial data reduction was performed using the standard FMOS {\sc fibre-pac} pipeline (Iwamuro et al. 2012), yielding extracted, calibrated 
spectra for each fibre. A correction for telluric absorption was made using stars observed in the field simultaneously with the galaxy targets. 
Because the FMOS OH suppression mask removes significant parts of each spectrum, 
rest-frame stacking is essential to derive clean and continuous spectra from these observations.
We constructed two independent composite spectra from the FMOS data, according to literature velocity dispersion measurements compiled by Smith et al. (2012a).
The low-$\sigma$ stack includes 32 galaxies with literature velocity dispersions in the range
 50\,$<$\,$\sigma$\,$<$\,100\,\kms, while the higher-$\sigma$ stack has 52 galaxies with 100\,$<$\,$\sigma$\,$<$\,300\,\kms.
By bootstrap resampling the input spectra used in the stacking process, we computed a set of 100 alternative realisations of the composite spectra, which 
are used to estimate the errors, including systematic variation among galaxies within the stack. 
The signal-to-noise ratio ($S/N$) estimated from the bootstraps is 100\,\AA$^{-1}$ (lower $\sigma$) and 200\,\AA$^{-1}$ (higher $\sigma$) at 1.14\,\mic.

\subsection{VLT SNELLS lens-candidate sample}

Very massive galaxies ($\sigma$\,$\ga$\,300\,\kms) are not well represented in the Coma sample. Since these objects are of particular interest, given the
evidence for non-standard IMFs in the most massive galaxies, we use a second dataset to sample this regime.

The observations for the high-mass sample were obtained for the ``SNELLS'' (SINFONI Nearby Elliptical Lens Locator Survey)  
project, a search for low-redshift strong-lensing galaxies among the massive early-type population
(Smith, Lucey \& Conroy 2015). The targets were drawn from the Six-degree Field (6dF) Peculiar Velocity Survey (Campbell et al. 2014), 
and the Sloan Digital Sky Survey (SDSS, Data Release 7, Abazajian et al. 2009), with 
primary selection  on velocity dispersion, with $\sigma$\,$>$\,300\,km\,s$^{-1}$ and redshift $z$\,$<$\,0.055. Galaxies with strong emission lines were
excluded.
We observed 26 candidate lenses in the J-band with the SINFONI integral-field spectrograph at the ESO 8.2\,m VLT UT4. 
The largest image scale was used, to provide a field-of-view 8$\times$8\,arcsec$^2$. The spectral resolution is $R$\,$\approx$\,2000. 
The observations were optimized for efficiency in finding background line emitters, with two 600\,sec exposures obtained per galaxy, offset spatially by 2.3\,arcsec. 

The spectrum for each galaxy was extracted from all spatial pixels with (absolute) total flux $>$50\,per cent of the maximum, after subtracting the two 
datacubes to remove the  background. This corresponds to an aperture of $\sim$1--2\,arcsec$^2$ in most cases. 
A telluric absorption correction was determined by fitting water and molecular oxygen transmission functions directly to the extracted spectrum, 
after dividing out a representative  passive galaxy model spectrum. 
As with the FMOS data, we combine the SNELLS spectra in the rest frame to derive a clean composite spectrum, and compute bootstrap realisations
 to quantify the errors. The derived $S/N$ is $\sim$350\,\AA$^{-1}$ at 1.14\,\mic.
The effective velocity dispersion, measured from the final J-band stack, is 297$\pm$7\,\kms, slightly smaller than expected given the SDSS/6dF measurements.

Among the SNELLS galaxies are two which were confirmed as strong lensing systems (SNL-1 and SNL-2) 
by Smith et al. (2015). 
The lensing analysis shows that these two galaxies have mass-to-light rations similar to what is expected for a MW-like IMF. 
Specifically, the mass-excess factors, relative to the Kroupa IMF, are
$\alpha$\,=\,1.20$\pm$0.13 and $\alpha$\,=\,0.94$\pm$0.17; hence bottom-heavy models are strongly disfavoured in these cases.

\begin{figure*}
\includegraphics[angle=270,width=155mm]{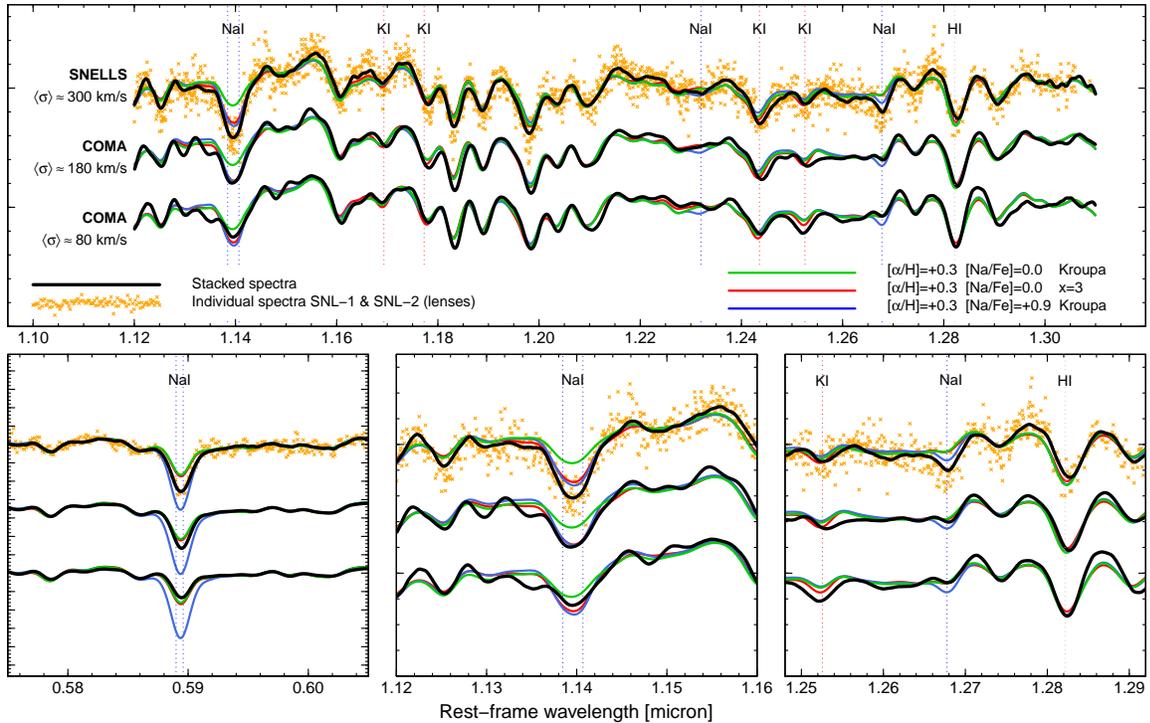}
\caption{Upper panel: composite J-band spectra for early-type galaxies in three bins of velocity dispersion. Lower left: stacked  SDSS and 6dF spectra
for the Na\,D region, using the same galaxies. Lower centre: zoom on the \naix\ doublet region in the J-band. Lower right: zoom on the weak Na\,{\sc i} line
at 1.27\,\mic. In each panel, the thick black lines show the stacked galaxy spectra, while the orange crosses are the individual flux points for two confirmed lenses SNL-1 and SNL-2 from Smith et al. (2015). Thin lines show the predictions from illustrative models from Conroy \& van Dokkum (2012a). The observed and model spectra 
have been smoothed to match the velocity dispersion of the SNELLS sample.}
\label{fig:allspectra}
\end{figure*}

\subsection{Optical spectra}

To aid in interpreting the \naix\ observations, we also use optical spectra to measure the Na\,D line at 0.58\,\mic.
The data are drawn from the 6dF survey (for most of the SNELLS galaxies) and  SDSS (for the Coma sample, and the remaining SNELLS 
galaxies). We created composite spectra and bootstrap realisations from the same galaxies as used in the FMOS and SNELLS stacks.

We note that the aperture sizes probed by the various datasets are not identical: the fibre sizes are 1.2\,arcsec, 3\,arcsec and 6.7\,arcsec in diameter, 
for FMOS, SDSS and 6dF respectively, while the effective aperture for the SNELLS spectra is 1--2\,arcsec. 
This is especially relevant when comparing the Na\,D measurements, taken from large apertures (SDSS and 6dF), against the J-band spectra obtained in small apertures from 
SNELLS and FMOS. 

Although the SNELLS spectra are derived from IFU data, our observing strategy is not suited to extracting spectra at larger aperture to match the SDSS/6dF spectra. 
However, we can estimate the likely effect of aperture mismatch using Na\,D measured in a range of aperture sizes in data cubes from
the CALIFA survey (S\'anchez et al. 2012). For 22 elliptical galaxies in the CALIFA second data release, 
the Na\,D line strength is reduced by 9 per cent on average, with a scatter of 5 per cent, when the aperture size is increased by a factor of five.
Correcting Na\,D upwards, to account for the smaller apertures in SNELLS
and FMOS, does not affect our conclusions, as we show in Section~\ref{sec:index}.

\section{Results}\label{sec:nai}

We present the results both through qualitative comparison of observed spectra against synthetic spectra, and through a more quantitative analysis using Lick-style line indices.
In both approaches, we use improved versions of the CvD models, as described by Conroy, Graves \& van Dokkum (2014).

\subsection{Spectral comparison}

\begin{figure*}
\includegraphics[angle=270,width=160mm]{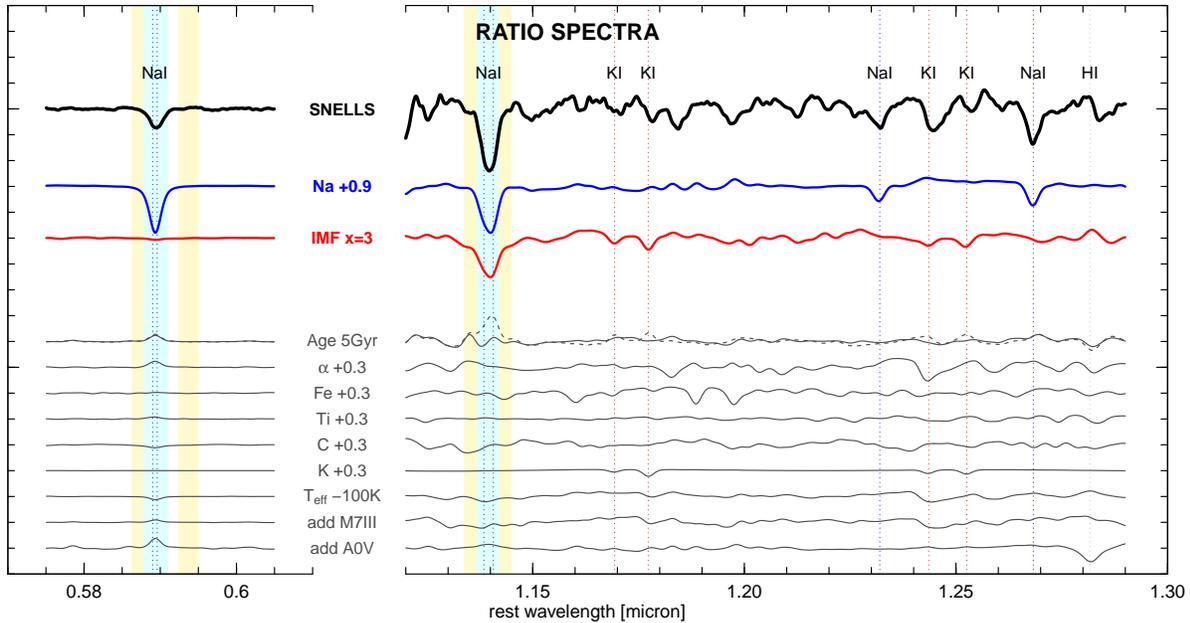}
\vskip -1mm
\caption{IMF and abundance effects on the J-band spectra and the Na\,D region.
The black line shows the ratio of the observed 
SNELLS stack versus a fiducial model (with [$\alpha$/H]\,=\,+0.3, [Fe/H]\,=\,[Na/H]\,=\,0, 13.5\,Gyr age and Kroupa IMF). The blue and red lines show the response spectra for 
sodium enhancement and IMF variation respectively. 
The effects of some other relevant parameters are shown below. 
{\rev The age response is shown both for the Kroupa IMF (solid) and for $x$\,=\,3 (dashed).}
Because Na\,D is intrinsically much stronger than the IR features, the ratio spectra in the Na\,D region have been scaled by a factor of 0.07 relative to the J-band for clearer
presentation.
Shaded regions indicate the feature and pseudo-continuum definitions of the \naix\ and Na\,D indices used in Section~\ref{sec:index}.
}
\label{fig:respillus}
\end{figure*}

The stacked spectra from the FMOS and SNELLS samples, together with the corresponding Na\,D stacks, are shown in Figure~\ref{fig:allspectra},
with illustrative model spectra for comparison. The FMOS stacks and the models have been smoothed to match the 
velocity dispersion of the SNELLS galaxies. 
Prominent spectral features in the J-band include the \naix\ doublet itself and the hydrogen Pa\,$\beta$ line at 1.28\,\mic, as well as many other 
heavily blended metal lines (see Rayner et al. 2008 for identifications).

As a fiducial comparison, we consider the 13.5\,Gyr model with Kroupa IMF and abundance pattern [$\alpha$/H]\,=\,+0.3; [Fe/H]\,=\,[Na/H]\,=\,0. 
This model clearly under-predicts both the \naix\ and the Na\,D absorption in the SNELLS sample. The discrepancy is smaller
in the FMOS stacks. The predicted \naix\ strength can be increased by adopting either a more bottom-heavy IMF or larger sodium 
abundances, but Na\,D is little affected by the IMF according to the CvD models. Hence some enhancement in [Na/H] is essential to match the SNELLS sample.
This is supported by the noticeable presence of a weak Na\,{\sc i} line at 1.27\,\mic, which only becomes visible in models with high [Na/H].
Taken individually, the two gravitational lenses, SNL-1 and SNL-2, both have spectra consistent with the SNELLS high-$\sigma$ composite, 
and in particular follow the behaviour of the stack in the Na\,D and \naix\ regions.

Figure~\ref{fig:respillus} shows the ratio between the observed SNELLS stack and the fiducial model, 
compared to `response spectra' showing the effect of some relevant stellar population parameters. 
The observed-vs-fiducial ratio spectrum confirms the excess absorption in the sodium lines, relative to the model. 
Only the [Na/H] and IMF responses have a strong impact on the \naix\ absorption. 
While either sodium enhancement or a bottom-heavy IMF could reproduce \naix, changes to these parameters lead to different predictions elsewhere, most
notably at Na\,D, but also in the weak sodium lines at 1.23\,\mic\  and 1.27\,\mic, and  the potassium doublets at 1.17\,\mic\ and 1.25\,\mic. 
Visual inspection of the \naix\ and Na\,D effects suggests a combination of sodium enhancement {\it and} 
bottom-heavy IMF is required for the high-$\sigma$ galaxies, if the combined effect of other parameters (age, etc) can be neglected.


\subsection{Line index comparison}\label{sec:index}

 
To summarize the behaviour of the strong sodium lines, we have measured \naix\ and Na\,D using  absorption indices.
For Na\,D, we use the Lick definition from Trager et al. (1998), while for \naix, we adopt a feature band of 1.1367--1.1420\,\mic, with pseudo-continua at 
1.1337--1.1367\,\mic\ and 1.1420--1.1450\,\mic\ (see Figure~\ref{fig:respillus}). All of the results are robust to small changes to these bandpasses.
For FMOS, the indices were measured after smoothing to match the velocity dispersion of the SNELLS stack. 
The index errors were derived from the bootstrap stacks, and hence explicitly include the galaxy-to-galaxy scatter within each sample.

Figure~\ref{fig:ixgrid} shows the measured indices, compared to a model grid spanning a range of [Na/H] and different IMFs. 
The indices were measured on the model spectra after smoothing to $\sigma$\,=\,300\,\kms\ to match the data. 
Note that the model grid is for [$\alpha$/H]\,=\,+0.3, which is appropriate for the SNELLS galaxies, but somewhat high for the FMOS stacks. 
(From optical spectra, the FMOS samples have average [Mg/Fe]\,=\,0.15 and 0.24.) 

The index--index diagram confirms and quantifies the results seen in the spectral comparisons:  \naix\  increases strongly with 
galaxy mass, deviating far from the predictions for solar [Na/H] and MW-like IMFs. Na\,D also increases strongly with $\sigma$, but not  
steeply enough t{o match a constant IMF model. Instead, the diagram implies a combination of both increasing [Na/H] and increasing dwarf-star content.
For the SNELLS stack, the grid suggests an IMF slope $x$\,$\approx$\,3.1 [where the Salpeter (1955) slope is 2.35] and [Na/H]\,$\approx$\,+0.5. 
{\rev Here, $x$ should be interpreted as a single-slope-equivalent value, since the detailed shape of the low-mass IMF is not constrained.}
Applying the CALIFA-derived aperture trend to correct Na\,D to the same aperture as used in the IR measurement, the inferred
IMF slope would be reduced slightly, to $x$\,$\approx$\,3.0, and the sodium abundance increased to [Na/H]\,$\approx$\,+0.6.
In the low-$\sigma$ FMOS stack, comparison with a model grid at lower [$\alpha$/H]  (not shown in the figure) leads to a derived 
[Na/H]\,$\approx$\,--0.3, and an IMF consistent with either Kroupa or Salpeter. 
The run of [Na/H] is thus similar to that found from stacked SDSS spectra by Conroy, Graves \& van Dokkum (2014), who
report [Na/H]\,=\,--0.24 at $\langle\sigma\rangle$\,=\,87\,\kms\ rising to [Na/H]\,=\,+0.43 at $\langle\sigma\rangle$\,=\,295\,\kms. 

Na\,D is a resonance line, subject to contamination by interstellar absorption.
Correcting for this would decrease the derived [Na/H] and increase the required $x$. 
However Jeong et al. (2013) studied ``Na\,D excess'' objects from SDSS  and concluded that the interstellar contribution is negligible for ordinary early-type galaxies.
{\rev Note that for steep IMFs, assuming ages younger than the nominal 13.5\,Gyr\ leads to weaker predicted absorption at \naix\ (and at Na\,D), 
and hence {\it larger} derived $x$ and [Na/H].}

Figure~\ref{fig:ixgrid} also shows the sodium indices for all of the individual SNELLS galaxies, highlighting the lenses SNL-1 and SNL-2. There is a wide 
spread in \naix\ among the sample, spanning the full range of model IMFs, at [Na/H]\,=\,0.3--0.5. As expected from the similarity of their spectra to the SNELLS composite, SNL-1 and SNL-2 have 
line strengths consistent with the properties of the stack, yielding [Na/H]\,$\approx$\,+0.4 and $x$\,=\,3.0--3.5. 
Thus despite having mass-to-light ratios consistent with a MW-like IMF, these galaxies have sodium indices characteristic of very steep IMF slopes, 
at least if interpreted using otherwise ``vanilla'' versions  of the CvD models (old, $\alpha$-enhanced, other abundances solar).

\section{Summary and discussion}\label{sec:disc}

The immediate result of this paper is that 
the \naix\ doublet is very strong in the most massive early-type galaxies. Synthetic models 
with Milky-Way-like IMF and solar [Na/H] fail to match the observed absorption. This behaviour is similar to that seen in the 
optical \naiop\ doublet, and suggests that the optical results are not due to the specific spectral ``environment'', e.g. 
contributions from the coincident Ti\,O band. 
Comparison with the CvD models suggests that reproducing the observed \naix\  requires high sodium abundances or bottom-heavy IMFs in massive galaxies.
We show that consistent results for  \naix\ and Na\,D can be obtained using a combination of {\it both} effects, 
with other parameters of the model held at plausible values.
For galaxies with $\sigma$\,$>$\,300\,\kms, we derive [Na/H]\,$\approx$\,+0.5 and an IMF with 
{\rev (single-slope-equivalent)} $x$\,$\approx$\,3. 
{\rev Two galaxies in our sample have strong-lensing masses compatible with an Milky-Way-like IMF, but their sodium indices imply 
steep IMFs as in the high-$\sigma$ stack. 
If the IMF really was a single power-law, the spectroscopic $M/L$ exceeds the lensing constraint by a factor of $\alpha$\,$\sim$\,4. 
Alternative IMF shapes, can reduce but not eliminate the discrepancy (e.g. fig. 5 of Smith 2014). For example the broken 
power-law assumed by La Barbera et al. (2013) would yield $\alpha$\,$\sim$\,2.5
}

We stress that the mass discrepancy applies specifically to results from the sodium indices, and {\it not} the general approach of fitting CvD 
models to constrain the IMF. For example, 
we have applied the full spectral fitting machinery of Conroy et al. (2014) to an extended SNELLS composite spectrum,
including wider optical coverage from 6dF/SDSS as well as SINFONI data in both J and H. The best-fitting model has a modest mass excess 
factor of $\alpha$\,$\approx$\,1.4, and [Na/H]\,$\approx$\,+0.7, and fits well over most of the  spectral range. 
This model has much more freedom than the simple  grid shown in Figure~\ref{fig:ixgrid}, with additional parameters describing the 
abundance pattern, two variable low-mass IMF slopes, and various nuisance effects. The greater flexibility allows the fit to 
reach closer to the observed line strengths than would be expected from the grid, given the recovered Salpeter-like IMF,  but the model
still under-predicts the measured \naix\ index by $\sim$5$\sigma$.

Our results suggest that the CvD models do not yet reproduce the strong  \naix\ in the most massive galaxies without violating other constraints 
(lensing masses, full-spectral fit). Such discrepancies in specific spectral features underline the advantage of combining many gravity-sensitive indicators
(e.g. Na\,{\sc i}, Ca\,{\sc ii}, Fe\,H) in deriving spectroscopic constraints on the IMF. They may also help to identify limitations
in the current models, and the corresponding opportunities for future improvements to the method.

\begin{figure}
\includegraphics[angle=270,width=84mm]{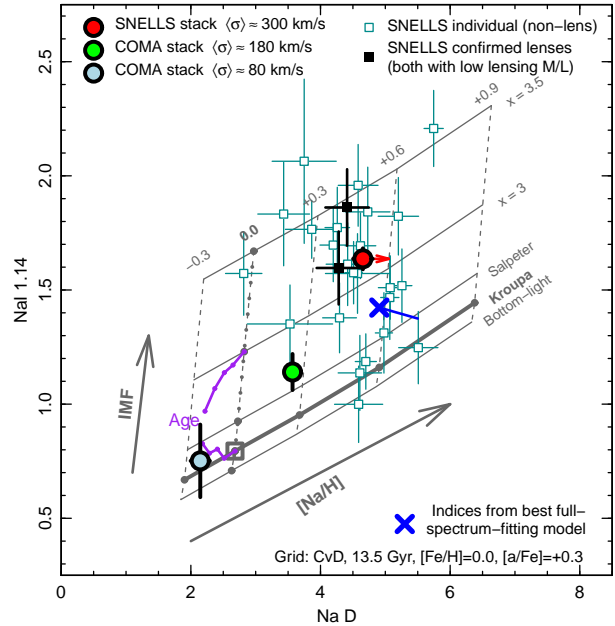}
\caption{Measured sodium indices compared to the grid computed from the CvD models for 13.5\,Gyr, [Fe/H]\,=\,0.0, [$\alpha$/H]\,=\,+0.3, with variable IMF and [Na/H]. The red arrow shows the estimated aperture correction for Na\,D in the SNELLS stack. The grey box is the ``fiducial'' model as in Figure~\ref{fig:allspectra}. 
Purple tracks show selected models at younger ages (down to 5\,Gyr).
The blue cross shows the indices
predicted by a full-spectrum fit to extended optical and J+H composite data for the SNELLS galaxies. The attached line segment shows how the indices 
are affected by extra parameters in the full-spectrum fit which are not accounted for in the two-parameter grid.}
\label{fig:ixgrid}
\end{figure}

\section*{Acknowledgements}

RJS was supported by the STFC Durham Astronomy Consolidated Grant 2014--2017 (ST/L00075X/1), PA by an STFC studentship, and DC by a Leverhulme Trust Emeritus Fellowship.
The data used here are available through the ESO science archive (programme ID 093.B-0193) and the Subaru archive (programme ID S14A-001).

{}

\label{lastpage}

\end{document}